\documentclass[final,5p,times,twocolumn]{elsarticle}
\usepackage{amsmath}
\usepackage{amssymb}
\usepackage{lineno}
\usepackage{wasysym}
\usepackage{color}
\usepackage{epstopdf}
\usepackage{mathrsfs}
\usepackage{siunitx}
\usepackage[latin1]{inputenc}
\usepackage{bbm}
\usepackage{graphicx}
\usepackage{graphics}
\usepackage{epsfig}
\usepackage{xcolor}
\usepackage{float}
\usepackage{ulem}
\usepackage[colorlinks,linkcolor=blue,anchorcolor=blue,citecolor=blue,urlcolor=blue]{hyperref}

\journal{Physics Letters A}

\begin{document}

\begin{frontmatter}

\title{Self-interfering dynamics in Bose-Einstein condensates with engineered dispersions}

\author{Jin Su$^{1,2}$, Hao Lyu$^{3}$, and Yongping Zhang$^{1}$\corref{*}}
\address{$^{1}$Department of Physics, Shanghai University, Shanghai 200444, China}
\address{$^{2}$ Department of Basic Medicine, Changzhi Medical College, Changzhi 046000, China}
\address{$^{3}$Quantum Systems Unit, Okinawa Institute of Science and Technology Graduate University, Onna, Okinawa 904-0495, Japan}

\ead{yongping11@t.shu.edu.cn}
\cortext[*]{Corresponding author}

\begin{abstract}
	
	 Optical lattice and spin-orbit coupling are typical experimental approaches to engineer dispersion. We reveal a self-interfering dynamics in a noninteracting Bose-Einstein condensate with the engineered dispersion by optical lattice or spin-orbit coupling. The self-interference results from the co-occupation of positive and negative effect mass regimes in the engineered dispersion. The physical origination of the self-interference is explained by the Wigner distribution function of the self-interfering wave-packet. We characterize detail features of the self-interference pattern.
\end{abstract}


\end{frontmatter}


\section{Introduction}

Dispersion relation plays an important role in wave-packet dynamics. Engineering dispersion becomes a significant method to control wave-packet dynamics.  
Effective mass is a fundamental concept to describe the motion of wave-packet in the engineered dispersion~\cite{Ashcroft}.  The effective mass is inversely proportional to the curvature of the dispersion. By steering the curvature, the effective mass can be negative. In such an engineered dispersion, the wave-packet acquires a negative effective mass. The negative effective mass can excite interesting phenomena in various systems such as optical systems, exciton-polariton condensates, and ultracold atoms~\cite{Cerda-Mendez,Mei,Scazza,Dusel,Mitchell}. Recently, Colas and Laussy have revealed a novel wave-packet in the engineered dispersion implemented in a noninteracting exciton-polariton condensates~\cite{Colas1}. This wave-packet as a single entity shows a counter-intuitive interfering pattern. Such self-interfering wave-packet originates from the co-occupation of dispersion regimes with both positive and negative effective mass~\cite{Colas1}. It is very different from a usual self-interference in which different parts of a single wave-packet overlap by external operations, such as expanding a wave-packet trapped in a toroidal trap inwards the origin~\cite{Toikka}.


In atomic Bose-Einstein condensates (BECs), optical lattice and spin-orbit coupling are two different ways for the engineering of dispersion.  The optical lattice opens band-gaps around Brillouin zone edges and gives rise to Bloch band-gap dispersion.  The optical lattice dispersion can be further managed by moving or shaking the lattice~\cite{Eiermann,Morsch,Parker}. With the optical lattice dispersion, various interesting physics relevant to the negative effective mass effect have been investigated, such as Bloch oscillation, gap soliton, and self-trapping~\cite{Dahan,Morsch2001,Eiermann2004,Anker}. In contrast, spin-orbit coupling manages the dispersion through steering kinetic energy~\cite{Goldman,Zhai,Zhang}. In atomic experiments, spin-orbit coupling is implemented by coupling two hyperfine states of atoms via a pair of Raman lasers~\cite{Lin,Wang,Cheuk,Ji}. Such Raman-induced spin-orbit coupling leads to a novel dispersion with a double-well structure in momentum space. Around the center of the momentum-double-well, the effective mass is negative.  A recent experiment has observed an exotic expansion dynamics induced by the spin-orbit-coupled dispersion with negative effective mass in a noninteracting BEC~\cite{Khamehchi}.  In the presence of spin-orbit-coupled dispersion and nonlinearity, interesting nonlinear dynamics have been predicted~\cite{Colas2018,Su}.

In this paper, we study self-interfering dynamics in a noninteracting BEC with optical lattice dispersion and spin-orbit-dispersion. Both dispersions have positive and negative effective mass regimes. This provides the possible existence of self-interfering dynamics according to Colas and Laussy~\cite{Colas1}. One of the fundamental features of BECs is coherence, which plays an essential role in matter wave dynamics. The coherence between two spatially separated BECs generates interference after they are merged~\cite{Andrews,Albiez}. The interference of spatially separated BECs has been experimentally observed in different BEC systems~\cite{Andrews,Albiez,Bongs,Anderson,Hadzibabic,Bloch}. For self-interference, there is no need to prepare two separated BECs. 
When the depth of optical lattice is very high, dynamics shall be confined in the lowest Bloch band. The tight-binding approximation, capturing the dynamics of the lowest band, offers a simple and efficient theoretical frame for relevant studies.  We construct a self-interfering wave-packet analytically thanks to the tight-binding approximation. We start from an initial Gaussian wave-packet which occupies both positive and negative effective mass regimes of the lowest band in momentum space. Its evolution in optical lattice generates self-interfering pattern.  The symmetry of the self-interfering pattern with respect to spatial coordinate can be broken for a nonzero quasimomentum of the initial Gaussian wave. Wigner distribution of the self-interfering wave-packet offers a clear picture for the origination of self-interfering pattern.  In the Wigner distribution, the wave-packet includes two different dominating quasimomenta coming from positive and negative effective mass regimes respectively. Mixing these two quasimomenta in space leads to interference.  The analytical results in optical lattice dispersion provide a useful guidance for revealing self-interfering dynamics in spin-orbit-coupled dispersion. We finally characterize the features of self-interference in a spin-orbit-coupled BEC.  

This paper is organized as follows. In Sec.~\ref{ol}, we construct a self-interfering wave-packet analytically in a noninteracting BEC loaded in an optical lattice. The interfering pattern is tunable by changing the initial quasimomentum. The physical origination of self-interference is uncovered from Wigner distribution. In Sec.~\ref{soc}, we study self-interfering dynamics in a spin-orbit-coupled BEC.  Section~\ref{conclusion} is the conclusion.

\section{Self-interfering dynamics in optical lattice}
\label{ol}

\begin{figure*}
\centering
\includegraphics[width=7in]{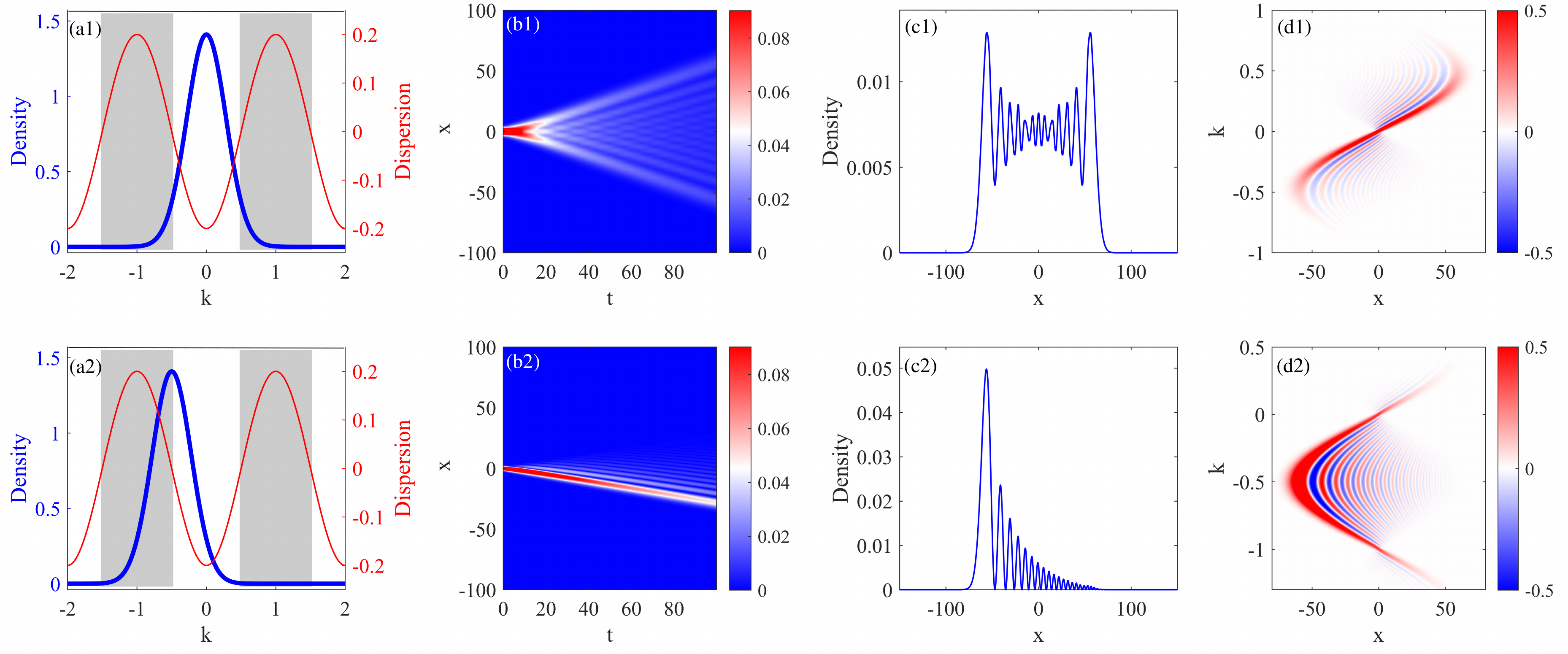}
\caption{A self-interference wave-packet in an optical lattice. (a1) Density distribution of an initial Gaussian wave-packet in momentum space (blue thick line) $|\tilde{\psi}(k,0)|^2$ with $k_0=0$.  The width of wave-packet is  $\sigma=2.5$. The red thin line are the dispersion relation in Eq.~(\ref{dispersion}) with $\omega_0=0.2$. The shadow areas indicate negative effective mass regimes. (b1) The time evolution of the density $n(x,t)=|\psi_L(x,t)|^2$ starting from the initial wave-packet in (a1).
(c1) A snapshot of the evolution in (b1) at  $t=100$. (d1) The Wigner distribution function $W(x,k)$ of the wave-packet shown in (c1). (a2)-(d2) are same as (a1)-(d1) but with initial quasimomentum $k_0=-0.5$. }
\label{Fig1}
\end{figure*}

Optical lattice is a fundamental experimental approach for quantum simulations. A noninteracting BEC loaded into an optical lattice is described by the Gross-Pitaevskii (GP) equation,
\begin{align}
i\hbar\frac{\partial}{\partial t}\psi(x,t)=H_L\psi(x,t),
\label{eq:OL}
\end{align}
with
\begin{align}
H_L=-\frac{1}{2}\frac{\partial^2}{\partial x^2}-V_L\cos(2x).
\end{align}
Here, $\psi(x,t)$ is the wave function and $V_{L}$ is the lattice depth. In the above dimensionless equation, the units of momentum, length, energy, and time are chosen as $\hbar k_L$, $1/k_L$, $\hbar^2k^2_L/(2m)$, and $2m/(\hbar k^2_L)$, respectively.  $m$ is the atom mass and $k_L$ is the wavelength of optical lattice lasers. 

The existence of optical lattice generates Bloch band-gap dispersion. In the tight-binding approximation, the dispersion of the lowest Bloch band becomes explicitly as~\cite{Bloch2005},
\begin{align}
\omega_L(k)=-\omega_0\cos(\pi k),
\label{dispersion}
\end{align}
 where, $\omega_L$ is spectrum, $\omega_0$ is the dispersion amplitude and relates to  $V_{L}$~\cite{Bloch2005} and $k$ is the quasimomentum.
 We assume that an initial wave-packet centers at $x=0$ and has a Gaussian shape,
\begin{align}
\psi(x,0)=\pi^{-\frac{1}{4}}\sigma^{-\frac{1}{2}}\exp\left( ik_0x-\frac{x^2}{2\sigma^{2}}\right) ,
\end{align} 
where $k_0$ is the initial quasimomentum and $\sigma$ characterizes the width. Its evolution in the lowest band is
\begin{align}
\psi(x,t)=\frac{1}{\sqrt{2\pi}}\int_{-\infty}^{+\infty}\tilde{\psi}(k,0)e^{i[kx-\omega_L(k) t]}dk,
\end{align}
with
\begin{align}
\label{initial}
\tilde{\psi}(k,0)&=\frac{1}{\sqrt{2\pi}}\int\psi(x,0)e^{-ikx}dx \notag \\ 
&=\pi^{-\frac{1}{4}}\sigma^{\frac{1}{2}}\exp\left[ -\frac{1}{2}\sigma^{2}(k-k_{0})^{2}\right] .
\end{align}
By using the Jacobi-Anger expansion~\cite{Gradstein},
\begin{align}
e^{i\omega_0t\cos(\pi k)}=\sum_{m=-\infty}^{\infty}i^{m}J_{m}(\omega_0t)e^{im\pi k},
\end{align}
the time-dependent wave function can be written as
\begin{align}
&\psi(x,t)=\sum_{m=-\infty}^{\infty}\psi_m(x,t), \notag \\
&
\psi_m(x,t)=c_{m}(t)\phi_{m}(x),
\label{eq:psi}
\end{align}
with 
\begin{align}
&c_{m}(t)=J_{m}(\omega_0 t)\exp\left[ik_{0}\left(x-m\pi \right)+\frac{im\pi}{2}\right] , \notag\\
&\phi_{m}(x)=\pi^{-\frac{1}{4}}\sigma^{-\frac{1}{2}}\exp\left[-\frac{\left(x-m \pi \right)^{2}}{2\sigma^{2}}\right].
\end{align}
Therefore, $\psi(x,t)$ can be regarded as the superposition of a series of Gaussian wave-packets $\phi_m(x)$, the locations of which are time-independent and are at $x=m\pi$. The superposition coefficients $c_m(t)$ depend on the initial momentum $k_0$ and the time $t$ via Bessel function $J_m(\omega_0 t)$.

The resulted analytical wave function is identified as a self-interfering wave-packet. We first study the case of $k_0=0$. The width of the initial wave-packet is chosen as $\sigma=2.5$.  The momentum-space density distribution $|\tilde{\psi}(k,0)|^2$ is shown in Fig.~\ref{Fig1}(a1).  The chosen width makes  the edges of the initial wave-packet extending into the negative effective mass regimes which are represented by the shadow areas in Fig.~\ref{Fig1}(a1). This initial state evolves according to Eq.~(\ref{eq:psi}), which is demonstrated in Fig.~\ref{Fig1}(b1). It is interesting to see that the wave-packet expands accompanied by the formation of interference fringes. Since there is no wave-packet splitting and recombination as the route of standard interference, the interference pattern in  Fig.~\ref{Fig1}(b1) is self-induced, called self-interference.   A snapshot at $t=100$ is shown in Fig.~\ref{Fig1}(c1).  From this figure, it is obvious that the wave-packet has very sharp edges. When $k_0=0$, it is known  from Eq.~(\ref{eq:psi}) that $\psi(x,t)=\psi(-x,t)$. Therefore, the wave-packet expands symmetrically with respect to the coordinate origin. The contrast of fringes increases significantly from the center to edges of the wave-packet.  The width of fringes around edges is wider than those around the center. There is no regular interference period. This makes not straightforward to see how the interference happens. 

In order to analyze the origination of self-interference, we calculate its Wigner distribution function.  Wigner distribution function is a representation of wave function in the phase space~\cite{Schleich}.  It is defined as
\begin{align}
W(x,k,t)=\frac{1}{2\pi}\int e^{-iuk}\psi^{*}\left(x-\frac{u}{2},t\right)\psi\left(x+\frac{u}{2},t\right)du.
\end{align}
By applying Eq.~(\ref{eq:psi}), we obtain
\begin{align}
W(x,k,t)&=\frac{1}{\pi}\exp\left(-\sigma^{2}k^{2}-2ik_{0}x\right)\sum_{m,n=-\infty}^{\infty}J_{m}(\omega_0t)J_{n}(\omega_0t)\notag\\
&\phantom{={}}\cdot \exp\left\lbrace  -\frac{1}{\sigma^{2}}\left[x-\frac{\pi}{2}(m+n)\right]^{2}+i\varphi\right\rbrace,
\end{align}
with
$$\varphi=\pi k(m-n)-\pi\left(\frac{1}{2}-k_{0}\right) \left(m+n\right).$$
 The Wigner distribution of the wave function at $t=100$ (corresponding to density shown in Fig.~\ref{Fig1}(c1)) is demonstrated in Fig.~\ref{Fig1}(d1). Red and blue color scales represent positive and negative values of $W(x,k,t)$ respectively. There is a symmetry $W(x,k,t)=W(-x,-k,t)$ since $\psi(x,t)$ is symmetric with respect to $x$.
The positive and negative value domains in the Wigner distribution in Fig.~\ref{Fig1}(d1) constitute a kind of interference structure in the phase space. The features of Wigner distribution are addressed as follows. 

(A)  For a fixed $x$, there are two dominating distributions at two different quasimomenta $k$ locating symmetrically around $k= -0.5$ for $x<0$ and around $k=0.5$ for $x>0$. The critical values $k=\pm 0.5$ are the location of inflection points in the dispersion of Eq.~(\ref{dispersion}). They are dividing lines between positive and negative effective mass regimes [also see Fig.~\ref{Fig1}(a1)]. Therefore, the one of two quasimomenta of dominating distributions comes from positive effective mass regime and the other comes from negative effective mass regime.  The superposition of wave-packets with these two quasimomenta generates to interference fringe at the fixed $x$. 

(B) The fringe period is inversely proportional to the difference between these two quasimomenta. From Fig.~\ref{Fig1}(d1), the difference is a largest value at $x=0$ and decreases with the increasing of $|x|$. This gives rise to the increasing of the fringe width as $|x|$ increases, as shown in Fig.~\ref{Fig1}(c1). At $x=\pm 63$, these two quasimomenta merge into one value of $k=-0.5$ for $x=-63$ and $k=0.5$ for $x=63$.  At $x=\pm 63 $ the difference between the two quasimomenta is so tiny that the interference fringe has a largest width [see Fig.~\ref{Fig1}(c1)].
Therefore, the sharp edges at $x=\pm 63$ come from the occupations of inflection points at $k=\pm 0.5$. From the dispersion, we know the velocity of a wave-packet at the inflection points is 
$v=\partial\omega_L(k)/\partial k|_{k=\pm 0.5}=\pm0.2\pi$. After $t=100$ evolution, the sharp edges are expected to appear at $x=vt\approx\pm63$.

(C) Furthermore, from Fig.~\ref{Fig1}(d1), the dominating occupation of quasimomentum in negative effective mass regime [i.e., $k<-0.5$ for $x<0$ and $k>0.5$ for $x>0$] 
increases with the increasing of $|x|$.  This causes that the fringe contrast around the edges is more obvious than around the center in Fig.~\ref{Fig1}(c1). The occupation of negative-effective-mass-quasimomentum affected by $|x|$ is due to the fact that the majority of the initial wave-packet lays in the positive effective mass regime and only the tails are in the negative effective mass regime. The quasimomentum is conserved during evolution. Therefore, the occupation of negative effective mass regime is fragile. 

\begin{figure}
	\centering
	\includegraphics[width=3.5in]{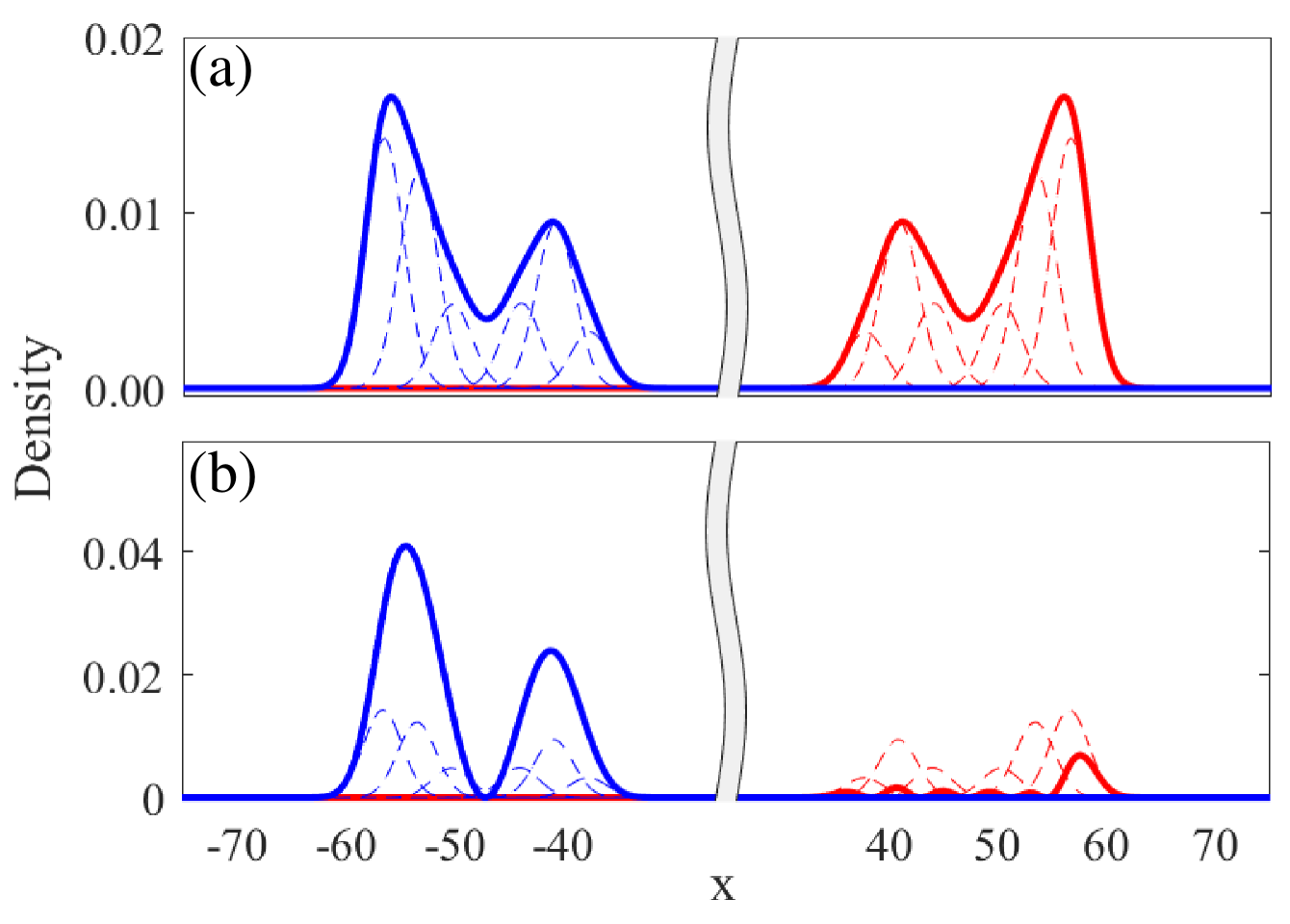}
	\caption{The total density $n(x,t)=|\psi(x,t)|$ from Eq.~(\ref{eq:psi})  (solid thick lines) and the density of Gaussian wave packets $n_m(x)=|\psi_m(x,t)|^2$ (dashed thin lines) at $t=100$.  In the positive coordinate regime $x>0$,  different dashed thin lines correspond to Gaussian waves with $m=12,13,\cdots,18$; in the $x<0$ regime, Gaussian waves are labeled by $-m=12,13,\cdots,18$. The distribution in (a) corresponds to the self-interference in Fig.~\ref{Fig1}(c1), and the distribution in (b) corresponds to that in Fig.~\ref{Fig1}(c2).
		} 
	\label{Fig2}
\end{figure}

At last, we want to emphasis that fringes in self-interference are due to the superposition of Gaussian wave-packets. Although the sum in Eq.~(\ref{eq:psi}) contains infinite modes, only finite modes make contributions to the atom density at a fixed $x$. To see how these Gaussian wave-packets interfere, we calculate density distribution around $x=\pm15\pi$ at $t=100$, which are close to the sharp edges. The wave function can be approximated as
\begin{align}
\psi(x,t)\approx \sum^{m_0+3}_{m=m_0-3}\psi_{m}(x,t),
\end{align}
where $m_0=15$ around  $x=15\pi$ and $m_0=-15$ around $x=-15\pi$. Other modes such as $\psi_{m_0\pm4}$ have much smaller amplitudes due to the very small $c_m$ so that they can be neglected.
 The total density $n(x)=|\psi(x,t)|^2$ and densities of Gaussian wave packets $n_m=|\psi_m|^2$ for $k_0=0$ are shown in Fig.~\ref{Fig2}(a). We can see that interference fringes are reproduced by the superposition of the Gaussian modes with $|m|=12,13,\cdots,18$.

In the following we study a nonzero initial quasimomentum $k_0=-0.5$. In experiments, such nonzero quasimomentum initial wave-packet can be prepared by accelerating the optical lattice~\cite{Eiermann2004}.  The initial wave-packet with a chosen width is shown in momentum space in Fig.~\ref{Fig1}(a2). This initial wave-packet distributes symmetrically around the inflection point at $k=-0.5$. Since $k_0\neq 0$, its evolution is not symmetric along $x=0$ from the Eq.~(\ref{eq:psi}) . The evolution is demonstrated in  Fig.~\ref{Fig1}(b2).  The wave-packet moves towards to the $-x$ direction.  This is because that the wave-packet locates in the dispersion with negative group velocity (i.e., $\partial\omega_L(k)/\partial k<0$). In the evolution, expansion is not obvious, but the appearance of interfering pattern is clear. Fig.~\ref{Fig1}(c2) shows the density distribution at $t=100$. It demonstrates a very interesting self-interference. The most outstanding is the interfering pattern is asymmetric with respect to $x=0$. With $k_0=-0.5$, in Eq.~(\ref{eq:psi}),  the superposition coefficients of Gaussian modes locating in $x>0$ are $c_m(t) $ and in $x<0$ are  $c_{-m}(t) $ ($m>0$). They are
\begin{align}
&c_{m}(t)=J_{m}(\omega_0t)\exp\left(-\frac{i x}{2}+im\pi\right), \notag \\ 
&c_{-m}(t)=J_{m}(\omega_0t)\exp\left(-\frac{ix}{2}\right). \notag
\end{align}
The phase $\exp(im\pi)$ in $c_{m}(t)$ makes the interference destructively in $x>0$. In Fig.~\ref{Fig2}(b), we show total density $n(x)=|\psi(x,t)|^2$  and dominating Gaussian modes $n_m=|\psi_m|^2$ around $x=\pm15\pi$ at $t=100$. The dominating Gaussian modes have same profiles around $x=\pm 15\pi$. However, the  phases between Gaussian modes around $x=15\pi$ make the total density very small. The other feature of self-interference in Fig.~\ref{Fig1}(c2) is that both fringe contrast and width decrease from left to right.  

The Wigner distribution of the wave function corresponding to that shown in Fig.~\ref{Fig1}(c2) is demonstrated in Fig.~\ref{Fig1}(d2). The distribution is symmetric along $k=-0.5$. This symmetry originates from the symmetric distribution of the initial wave-packet along $k=-0.5$. The dominating occupation is the outer-red domain. For a fixed $x$, there are two quasimomenta coming from the negative and positive effective mass regimes in the outer-red domain. Superposition of these two quasimomentum components leads to interference structure in Fig.~\ref{Fig1}(c2). The difference between these two quasimomenta decreases from right to left, and they merge together into $k=-0.5$ at $x=-63$.  This results in attenuation of fringe width from left to right.  The distribution values of outer-red domain in $x>0$ are obviously smaller than these in $x<0$. Therefore, the fringe contrast is large in $x<0$. Specifically, the most population in Fig.~\ref{Fig1}(d2) concentrates around $k=-0.5$, leading to a large density peak around $x=-63$ in  Fig.~\ref{Fig1}(c2).

\begin{figure*}[t]
	\centering
	\includegraphics[width=7in]{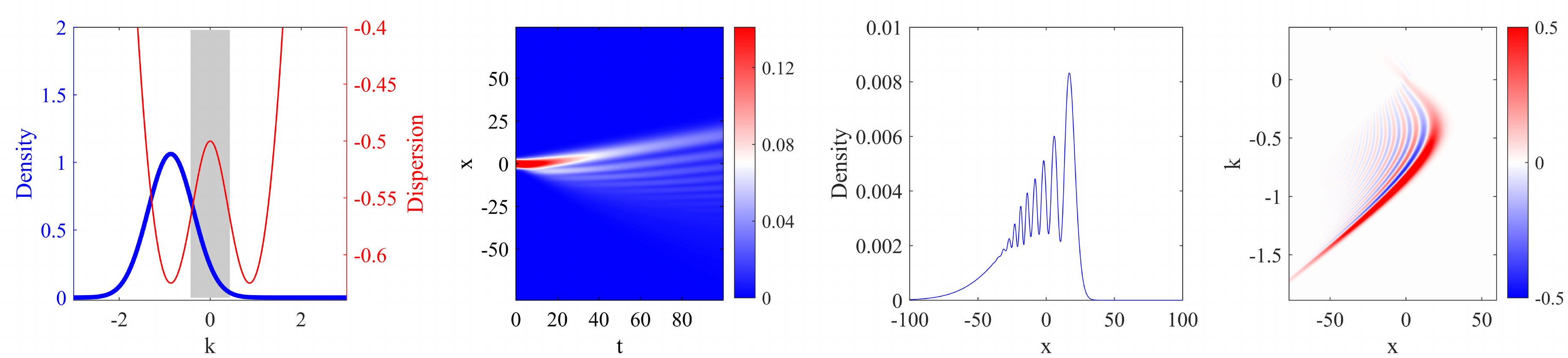}
	\caption{A self-interfering wave-packet in a spin-orbit coupled BEC. (a)  Density distribution of the initial wave-packet in momentum space (blue thick line) and the spin-orbit-coupled dispersion in Eq.~(\ref{socdispersion}) (red thin line).
  (b) The evolution of the density $n(x,t)=|\psi_{\mathrm{soc}}(x,t)|^2$ as a function of time $t$. (c) Snapshot of the time evolution at $t=100$. (d) Wigner distribution function $W(x,k)$ of the wave-packet at $t=100$. The parameters are $\sigma=2$, $\gamma=1$, and $\Omega=1$.}
	\label{Fig3}
\end{figure*}

We want to emphasis that the Wigner function provides a very useful approach to understand the self-interference in optical lattice. From the Wigner function, we know that the co-occupation of both positive and negative effective mass regimes responses for the self-interference. The self-interfering wave-packet in Eq.~(\ref{eq:psi}) is the superposition of Gaussian wave-packets centering at each unit of optical lattice.  These Gaussian wave-packets may be considered as lowest eigenstates of each lattice unit. Therefore, one may think that the lattice structure is important for the self-interference.  Optical lattice is the system in which  lattice structure is in coexistence with the dispersion processing both positive and negative effective mass regimes.  This makes it not straightforward to see the origination of  self-interference laying at the engineered dispersion. In the following section, we reveal the existence of self-interference in a noninteracting spin-orbit-coupled BEC. The spin-orbit-coupled BEC is a homogeneous system, there is no external lattice structure. The study of  self-interfering dynamics in this system enables a straightforward understanding of its origination.

\section{Self-interfering dynamics in a spin-orbit coupled BEC}
\label{soc}

The spin-orbit-coupled Hamiltonian is ~\cite{Khamehchi,Colas2018}
\begin{align}
H_{\mathrm{SOC}}=-\frac{1}{2}\frac{\partial^2}{\partial x^2}-i\gamma\frac{\partial}{\partial x}\sigma_z+\frac{\Omega}{2}\sigma_x,
\label{eq:soc}
\end{align}  
where  the first term on the right side is  the kinetic energy, the second term is the spin-orbit coupling with $\gamma$ being coupling strength, and  $\Omega$ is the Rabi frequency which depends on the laser intensity.  $\sigma_{x,z}$ are Pauli matrices.  In the above Hamiltonian, the units of momentum, length, energy, and time are $\hbar k_{\mathrm{Ram}}$, $1/k_{\mathrm{Ram}}$, $\hbar^2k^2_{\mathrm{Ram}}/m$, and $m/(\hbar k^2_{\mathrm{Ram}})$, respectively. Here, $k_{\mathrm{Ram}}=2\pi/\lambda_{\mathrm{Ram}}$ is the wavenumber of the Raman lasers with $\lambda_{\mathrm{Ram}}$ being the corresponding wave length.

The dispersion of Eq.~(\ref{eq:soc}) has two branches with an energy gap depending on the Rabi frequency~\cite{Zhang}. We assume the energy gap is so large that dynamics are confined in the lower branch and we neglect the upper branch. The dispersion of the lower branch is 
\begin{align}
\omega_{\mathrm{SOC}}(k)=\frac{k^{2}}{2}-\sqrt{\gamma^2 k^{2}+\frac{\Omega^{2}}{4}}.
\label{socdispersion}
\end{align}
If $\Omega<2\gamma^2$, the dispersion has two parabolas located at $\pm k_m$ with $k_m=\gamma\sqrt{1-\Omega^2/(4\gamma^4)}$. This dispersion is shown by red line in Fig.~\ref{Fig3}(a).  A negative-mass regime emerges around $k=0$, which is represented by the shaded area. 
From $\partial^2\omega_{\mathrm{SOC}}(k)/\partial k^2=0$, we obtain the locations of inflection points $\pm k_n$, which separate positive and negative effective mass regimes. 
\begin{align}
k_n=\frac{1}{2\gamma}\sqrt{\left(2\gamma^2\Omega^2\right)^{2/3}-\Omega^2}.\notag
\end{align}
The group velocity can be calculated from the dispersion, $v_{\mathrm{SOC}}(k)=\partial\omega_{\mathrm{SOC}}(k)/\partial k$. We find that $v_{\mathrm{SOC}}(k)$ has a local maximum at $-k_n$ and a local minimum at $k_n$. The eigenstate corresponding to the lower branch dispersion is 
\begin{align}
\phi_k=\vec{s}(k) \exp(ikx),
\end{align}
with
\begin{align}
\vec{s}(k)=\frac{1}{\sqrt{\Omega^2+4\gamma^2(-k+\gamma)^2}}\left[ 
\begin{array}{c} \Omega \\ 2\gamma(k-\gamma) \end{array} \right],
\end{align}

We construct an initial wave-packet as
\begin{align}
\psi_{\mathrm{SOC}}(x,0)=\pi^{-1/4} \sigma^{-1/2}\exp\left(  -\frac{x^2}{2\sigma^{2}}\right) \phi_{-k_m},
\end{align}
Here, $ \phi_{-k_m}$ is the ground eigenstate of $H_{\mathrm{SOC}}$ with the quasimomentum $-k_m$.  The initial wave-packet is the ground eigenstates modulated by a Gaussian function.  In spin-orbit-coupled experiment, such initial state can be prepared by adiabatically dressing a harmonic trapped BEC with Raman lasers.  In momentum space, the initial wave-packet is Fourier transformed as
 \begin{align}
 \tilde{\psi}_{\mathrm{SOC}}(k,0)=\pi^{-1/4} \sigma^{1/2}\vec{s}(k_m)\exp\left[ -\frac{1}{2}\sigma^2(k+k_m)^2\right] . \notag
 \end{align}
With a chosen width $\sigma$, the distribution $ | \tilde{\psi}_{\mathrm{soc}}(k,0) |^2$ is demonstrated in Fig.~\ref{Fig3}(a). A part of the initial wave-packet  on the right side  enters into the negative effective mass regime, while the remaining in the positive effective mass regime. 

The time evolution of such initial wave-packet in the spin-orbit-coupled dispersion of Eq.~(\ref{socdispersion}) is
\begin{align}
\psi_{\mathrm{SOC}}(x,t)=\int_{-\infty}^{\infty} \tilde{\psi}_{\mathrm{SOC}}(k,0) e^{i[kx-\omega_{\mathrm{SOC}}(k)t]}dk,
\end{align}
 Figure~\ref{Fig3}(b) shows the evolution result of the density $|\psi_{\mathrm{SOC}}(x,t)|^2$.  The evolution presents a perfect self-interference.  This interference structure is similar as that in Fig.~\ref{Fig1}(b2).  The  density distribution at $t=100$ is shown in Fig.~\ref{Fig3}(c).  Interference fringes are very obvious in the $x>0$ region, accompanying with a sharp and big edge peak.  There is no inference fringe in the left tail of wave-packet.  The Wigner function of this wave-packet is calculated numerically and demonstrated in Fig.~\ref{Fig3}(d).  Similar to the self-interference in optical lattice, the features of interference in Fig.~\ref{Fig3}(c) can be fully explained by the Wigner function.  The outer-red domain dominates in the Wigner function, which gives a main contribution to the interference pattern of Fig.~\ref{Fig3}(c).  For a fixed $x$, there are two major occupations with different quasimomenta in the outer-red domain.  The turning point of the outer-red domain lays at  $k=-k_n, x=22.5$, which means that the big edge density peak on the right side in Fig.~\ref{Fig3}(c) comes from the occupation of the dispersion at $k=-k_n$. This can be further checked that the edge peak position is $x=22.5$ corresponding to $v_{\mathrm{SOC}}(k)t$. For a very negative $x$, there is only one occupation in the outer-red domain, leading to no interference on the very left side of wave-packet. This is because that the left tail comes from the occupation of the dispersion on the left side [see Fig.~\ref{Fig3}(a)], and there is no negative effective regime on this part of dispersion.  Since the velocity of this part of dispersion is negative, the left tail of wave-packet expands towards negative direction without appearance of interference.

\section{Conclusion}
\label{conclusion}

We study the self-interfering dynamics of a noninteracting BEC in optical lattice or with spin-orbit coupling. Both the optical lattice and spin-orbit coupling can engineer dispersion relations. Self-interference in the engineered dispersion originates from the co-occupation of  positive and negative effective regimes in dispersion. The separation between these two regimes is group velocity maximum. Atoms occupying the separation shall have a fastest movement to form the edges of a self-interfering wave-packet.  Around the separation, the engineered dispersion always has a pair of equal group velocities, one of them comes from the positive effective mass regime and the other from the negative effective mass regime. Atoms occupying these two dispersion points always move with the same velocity.
Therefore they can overlap coherently in space to generate interference.


	\textbf{Acknowledgements}
	This work is supported by National Natural Science Foundation of China with Grants Nos.~11974235 and 11774219.

\end{document}